\newcommand\lsim{\mathrel{\rlap{\lower4pt\hbox{\hskip1pt$\sim$}}
    \raise1pt\hbox{$<$}}}
\newcommand\gsim{\mathrel{\rlap{\lower4pt\hbox{\hskip1pt$\sim$}}
    \raise1pt\hbox{$>$}}}

\documentstyle[prl,aps,epsf,floats,twocolumn]{revtex}

\begin{document}
\twocolumn[\hsize\textwidth\columnwidth\hsize\csname @twocolumnfalse\endcsname
\title{Cosmic string loops and large-scale structure}

\author{P.P.~Avelino$^{1}$,
E.P.S.~Shellard$^{2}$, J.H.P.~Wu$^{2}$ and B. Allen$^3$}
\address{$^{1}$Centro de Astrof{\' \i}sica, Universidade do Porto, Rua das 
Estrelas s/n,
4150 Porto, Portugal}
\address{$^{2}$Department of Applied Mathematics and Theoretical Physics,
University of Cambridge,\\
Silver Street, Cambridge CB3 9EW, U.K.}
\address{$^{3}$University of Wisconsin---Milwaukee, U.S.A.}

\maketitle

\begin{abstract}
We investigate the contribution made by small loops from a 
cosmic string network as seeds for large-scale structure
formation.
We show that
cosmic string loops are highly correlated with the long-string network
on large scales and therefore
contribute significantly to the power spectrum of density perturbations
if the average loop lifetime is comparable to or above one Hubble time.
This effect further improves the large-scale bias problem
previously identified in earlier studies of cosmic string models.
\end{abstract}
\vskip .2in
]



\subsection{\bf Introduction}
\label{intro}

Quantitative predictions for the large-scale structure induced 
by cosmic strings have taken some time to crystallise as the 
understanding of cosmic string physics has improved \cite{VilShe}.  
In particular,
the role of small loops produced by the string network has evolved from 
a potential one-to-one correspondence between loops and cosmological objects
\cite{Vil} through to a completely subsidiary role relative to the wakes
swept out by long strings \cite{SilZel}. This dethronement of loops
was a result of numerical studies which showed that the average loop 
size $\bar\ell = \alpha t$ was much smaller than the horizon, 
$\bar\ell <\!< d_{\rm H}$ \cite{AS1,BenBou2};
they might even be as small as the lengthscale set by gravitational 
backreaction $\alpha \sim 10^{-4}$, a value appropriate for GUT-scale strings 
\cite{VilShe}.  Add the high ballistic 
loop velocities observed $\bar v\approx c/\sqrt{2}$ 
and it was not 
surprising that these tiny loops have been assumed to be more or less
uniformly  distributed and hence a negligible source relative to the long 
string network \cite{Ste}.  Nevertheless, small loops always make up a 
significant 
fraction of the total string energy density at any one time and, as we
demonstrate here, loop-induced inhomogeneities are considerable 
if their lifetime
is not much smaller than the Hubble time.  By properly 
incorporating these loop perturbations, we show that their contribution 
relative to the long string wakes is almost comparable and also highly 
correlated with these wakes.

The context for this work is a major programme of structure formation 
simulations seeded by high resolution cosmic string networks with  very 
large 
dynamic ranges \cite{AveShe2,AveShe4,AveShe3}.
This work demonstrated
that for open or $\Lambda$ models with 
$\Gamma\,=\,\Omega h\,=\,0.1$--$0.2$ and a cold dark 
matter (CDM) background,
the linear density fluctuation power
spectrum has both an amplitude at $8 h^{-1}$Mpc, $\sigma_8$, and an 
overall shape which are consistent within uncertainties with 
those currently inferred from galaxy surveys.
This result has also 
been confirmed using semi-analytical phenomenological models which incorporated 
some of the main features of long string networks
\cite{AveCal1,BRA}.

In this letter we investigate the contribution of cosmic string loops to the 
linear power spectrum of cosmic string induced density perturbations.
This component has been ignored and excluded in previous work,
due to both the computational difficulties and assumptions about the homogeneity
of the loop distribution.  To this end
we first perform very high resolution numerical simulations of a
cosmic string network with a dynamic range extending from well before the
radiation-matter transition through to deep into the matter era.
We then use this network as 
a source for density perturbations (as described in \cite{AveShe2,AveShe3})
taking into account
the large-scale power contributed from cosmic string loops. 
This is done 
by modelling cosmic string loops smaller than a fixed fraction of the horizon 
size as relativistic point masses.
The effects of the evaporation of these loops into 
gravitational waves and the damping of loop motion due to expansion 
are also included.   Note, however, that this 
should be clearly distinguished from recent work \cite{ConHin},
which attempts to incorporate network decay products in the power spectrum
of an additional fluid (with a variety of possible equations of state).
This does not appear to properly account for the phase correlation between
long strings and moving loops.

Unless otherwise indicated,
we use $h=0.7$, $\Omega_{\rm m}=1$ and $\Omega_\Lambda=0$
in the results presented here.
A verified accurate rescaling scheme for the resulting power spectrum
with different choices of $h$, $\Omega_{\rm m}$ and $\Omega_\Lambda$
is straightforward and described in
ref.\cite{AveShe2,AveShe3,AveCal1,AveCar}.

\subsection{\bf Cosmic string and loop evolution}
\label{stringsource}

\noindent The Nambu equations of motion  for cosmic strings in an expanding
universe can be averaged to yield:
\begin{equation}
  {{d \rho_\infty} \over dt}+2H(1+{\langle v^2 \rangle}) \rho_\infty= - X_L, 
  \label{rhotwo}
\end{equation}
where
$\rho_\infty$ is the long string energy,
$t$ the physical time,
$H=\dot{a}/a$  the Hubble parameter, $a(t)$ the scale factor,
$\langle v^2 \rangle$ the mean square velocity of strings,
and 
$X_L$ the transfer rate of energy density from long strings into loops.
In the scaling regime the 
long string energy density should scale with the background
energy density  evolving as
\begin{equation}
  {{d \rho_\infty} \over dt} = -2{{\rho_\infty} \over t}.
  \label{rhoone}
\end{equation}
Substituting this into (\ref{rhotwo}) to eliminate $d \rho_\infty/dt$ gives
\begin{equation}
  {{t X_L} \over \rho_\infty}=
  \left\{
    \begin{array}{ll}
      (1-{\langle v^2_{\rm r} \rangle}) \sim 0.6
      & {\rm in\ radiation\ era},\\
      {2  \over 3} (1-2 {\langle v^2_{\rm m} \rangle}) \sim 0.2
      & {\rm in\ matter\ era},
    \end{array}
  \right.
  \label{one}
\end{equation}
where ${\langle v^2_{\rm r} \rangle} \gsim {\langle v^2_{\rm m} \rangle} \sim 0.6$
\cite{AS1,BenBou2}.
Both (\ref{rhoone}) and (\ref{one}) provide a check for the scaling behavior of
long strings and loops in the cosmic string network simulations.

We know that the loops produced by a cosmic string network
will decay into gravitational 
radiation, with a roughly constant decay rate $\Gamma G \mu^2$,
where $\mu$ is the string linear energy density.
Typically $\Gamma=50-100$ with
an average $\langle \Gamma \rangle \sim 65$\cite{SchQua,AllShe}. 
Now, if we assume the loop production to be `monochromatic' so that all loops 
formed at the same time will have the same mass,
we can write the initial rest mass of a loop formed at time $t_*$ as 
\begin{equation}
  M_L^*=\alpha\mu t_* \equiv f \Gamma G \mu^2 t_*\ .
  \label{inimass}
\end{equation}
Here, the parameter $f = \alpha/\Gamma G \mu$ is 
expected to be of order unity if the size of the loops formed at the time $t$ 
is determined by gravitational radiation back-reaction,
which smoothes strings on scales smaller than $\Gamma G \mu t$. 
With the decay rate introduced earlier,
we have the rest mass of a loop formed at time $t_*$ evolving as
\begin{equation}
  M_L(t_*, t)= M_L^* W(t_*, t)\ ,
  \label{massdecay}
\end{equation}
where  
\begin{equation}
  W(t_*, t)=
  \left\{
    \begin{array}{ll}
      1- {t-t_* \over \tau(t_*)} &
      {\rm for \ } t_* \le t \le t_*+\tau(t_*) \\
      0 & {\rm otherwise}
    \end{array}
  \right.\ .
\label{four}
\end{equation}
Here $\tau(t_*)\approx ft_*$ is the lifetime of loops produced at time $t_*$
($f=2$, $3$ implies the decay occurs in one horizon time in the radiation 
and matter eras
respectively).
The evolution of the loop energy density is then given by:
\begin{eqnarray}
    \rho_L (t)
    & = &
    \int_0^{t} X_L(t') {\left[{{a(t')} \over {a(t)}}\right]}^3 W(t', t) dt'
    \nonumber
    \\
    & \propto &
    \left\{
      \begin{array}{ll}
        f/t^2        & {\rm for\ } f \ll 1,\\
        \sqrt{f}/t^2 & {\rm for\ } f \gg 1 {\rm \ (radiation\ era),}\\
        (\ln{f})/t^2 & {\rm for\ } f \gg 1 {\rm \ (matter\ era),}
      \end{array}  
    \right.
    \label{rhol}
\end{eqnarray}
where we have used the scaling behaviour
(\ref{rhoone}) and (\ref{one}).
Consequently, the scaling of the power spectrum induced by loops in $f$
should interpolate between
$f^2$ and $f$ (radiation era) or $(\ln{f})^2$ (matter era).
We notice in (\ref{rhol}) that we have ignored the effect of loop velocity
redshifting
due to the expansion of the Universe, which causes a change in the
effective mass.
Because loops are formed with relativistic velocities,
we expect this damping mechanism to have the strongest effect for
$f \gg 1$, but to be negligible for  $f \ll 1$.

If a loop formed at time $t_*$ has an initial physical 
velocity ${\bf v}_*$, its trajectory in physical space
accounting for the expansion of the Universe is then given by:
\begin{equation}
  {\bf x}(t)
  = {\bf x}(t_*) +
  a(t) \int_{t_*}^t {{\bf A} \over {\sqrt{a(t')^2 + A^2}}} {dt' \over a(t')}
  \label{five}
\end{equation}
for $t \ge t_*$,
where
${\bf A}=\gamma_* {\bf v}_* a_*$, $A=|{\bf A}|$ and 
$\gamma_*=(1-|{\bf v}_*|^2)^{1/2}$.
Here we have neglected the acceleration of loops
due to the momentum carried away by the gravitational radiation, the 
so-called `rocket effect'.
A numerical calculation for several asymmetric loops shows that
the rate of momentum radiation from an oscillating loop is
\begin{equation}
|{\dot {\bf P}}|=\Gamma_P G \mu^2\ ,
\end{equation}
where $\Gamma_P \sim 10$\cite{VacVil}.
Combining with (\ref{massdecay}),
one can show that this rocket effect will become important 
only when:
\begin{equation}
  {t \over t_*} \gsim
  1+ {f \over 1+ \Gamma_P/(\Gamma \gamma v)} \ ,
\end{equation}
which affects only the final stages of the loop lifetime
as long as $\Gamma_P/(\Gamma \gamma v) < 1$, or equivalently $v \gsim 0.15c$.
For a typical $v_*\sim c/\sqrt2$,
one requires a loop lifetime $\gsim 43t_*$ in the radiation era
and $\gsim 17t_*$ in the matter era to redshift down to
this critical velocity according to (\ref{five}).
Since the values of $f$ we explore here are of order unity,
it is a reasonable approximation to neglect
the transfer of momentum due to gravitational radiation.

\subsection{\bf Results and discussion}
\label{biaf}

\begin{figure}
\centering 
\leavevmode\epsfxsize=8cm \epsfbox{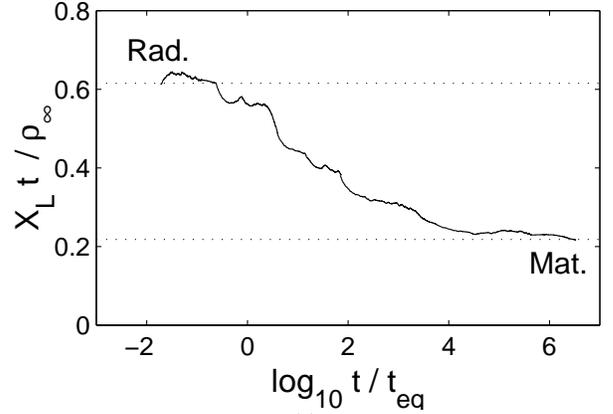}\\ 
 \caption[]
  {Evolution of $X_L(t)$.
    The dotted lines are the
    asymptotic values in the radiation and the matter eras.}
  \label{figure1}
\end{figure}

We first perform string simulations with a string sampling spacing
$1/1000$ of the simulation box sizes.
The dynamic ranges cover from $0.05$ to $300$ $\eta_{\rm eq}$,
where $\eta_{\rm eq}$ is the conformal time
at radiation-matter energy density equality.
We then perform the structure formation simulations
with box sizes ranging from $20$--$120h^{-1}$Mpc,
and a resolution of $128^3$--$512^3$.
Figure \ref{figure1} shows the evolution of $X_L$.
We can see that the expected amount of energy
was converted into loops in our simulations so that
$X_L$ has the correct asymptotic behavior given by (\ref{one}).
However,
the typical loop-size
(and consequently their lifetime) 
does not approach scaling so rapidly and is therefore larger than 
physically expected for most of the duration in the simulations
\cite{AS1,BenBou2}. 
To overcome this problem
we rescale the loop lifetime according to equation (\ref{inimass}).
Thus the uncertainty in the average mass and therefore the lifetime of loops
formed at a given time
is quantified by the choice of the parameter $f$.
The initial rms velocity of loops observed from the simulations
is $\langle v^2_*\rangle^{1/2} \gsim 0.7c$ throughout all the regimes.

\begin{figure}
\centering 
\leavevmode\epsfxsize=8cm \epsfbox{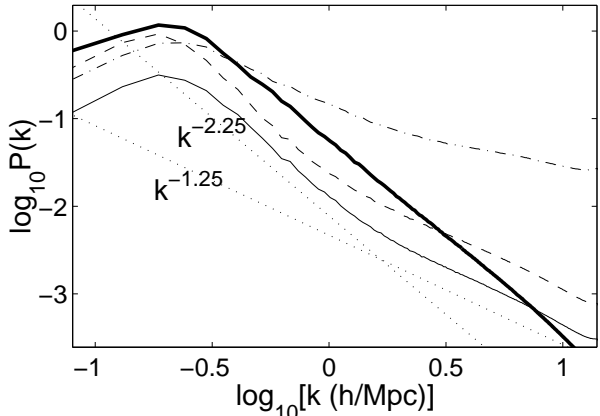}\\ 
 \caption[]
 {Small dynamic range power spectra of density perturbations seeded
   by long strings (thick solid),
   by loops with initial velocities $v_*$ switched to zero (dot-dashed), and
   by loops with $v_*$ determined by string network evolution
   (dashed and thin solid).
   The thin solid line includes the effect of
   gravitational decay of the loop energy,
   while the other two loop lines don't but with loops removed 
   after a period of time $\tau_*=t_*$.
   }
 \label{figure2}
\end{figure}

Figure \ref{figure2} shows the power spectrum of density perturbations
induced by long strings and by cosmic string loops for $f=1$ for a small 
dynamic range from $2.5$ to $5\eta_{\rm eq}$.
We can see that
when compared with the spectrum induced by static loops (dot dashed),
the amplitude of small-scale perturbations induced by moving loops (dashed)
is clearly reduced by their motion.  However, their large-scale power is 
higher because of the dependence of the gravitational interaction 
on the loop velocities,
especially when they are relativistic.
We also see that the gravitational decay of loop energy (thin solid)
damps the overall amplitude of the power spectrum (dashed)
by about a factor of 3.
We notice that between the long-string correlation scale
$k_\xi \approx 20/\eta \cite{AveShe2,AveShe4,AveShe3}$
and the scale $k_{\rm L}\approx 10 k_\xi$,
the slope of the loop spectrum (thin solid) is exactly the same as that of
the long-string spectrum $n \approx -2.25$ \cite{AveShe3}.
We believe that this close correspondence is due to copious loop
production being strongly correlated with long 
string intercommuting events and the collapse of highly curved long string
regions \cite{AS1}, that is, near the strongest long string perturbations.
Moreover, these correlations persist in time with the subsequent 
motion of loops and long strings lying preferentially in the same directions,
a phenomenon which  has been verified by observing animations of
string network evolution.
These correlations between loops and long strings, however, have a lower 
cutoff represented by the mean loop spacing $d_{\rm L}\sim k_{\rm L}^{-1}$.
Below $d_{\rm L}$, the effects of individual filaments swept out by moving 
loops can be identified. In terms of the power spectrum,  for $k<k_{\rm L}$
the loops are 
strongly correlated with the 
long strings and therefore reinforce the wake-like perturbations, while 
for $k>k_{\rm L}$ their filamentary perturbations increase the spectral 
index by about one to $n \approx -1.25$; this change is expected on geometrical 
grounds.

\begin{figure}[t]
\centering 
\leavevmode\epsfxsize=8cm \epsfbox{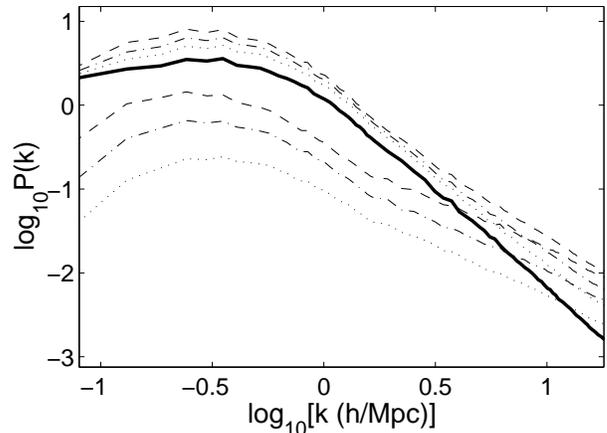}\\ 
 \caption[]
 {The lower set of 3 lines are ${\cal P}_{\rm L}(k)$ for 
  $f=0.5$ (dotted), $1$ (dot-dashed) and $2$ (dashed).
  ${\cal P}_\infty (k)$ is plotted as a solid line.
  The upper set of lines are ${\cal P}_{\rm tot}(k)$
  with corresponding line styles and $f$ values
  to the lower set of lines.
  }
  \label{figure3}
\end{figure}
\begin{figure}[t]
\centering 
\leavevmode\epsfxsize=8cm \epsfbox{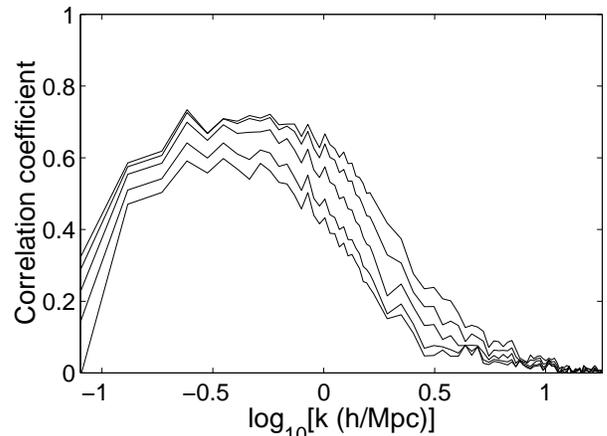}\\ 
 \caption[]
 {The correlation coefficient between the long-string and
   loop induced perturbations,
   with $f=0.5, 1, 2, 4, 6$ (downwards).
   }
  \label{figure5}
\end{figure}
In figure \ref{figure3} we plot the power spectra of
density perturbations seeded
by long strings ${\cal P}_\infty (k)$,
by small loops ${\cal P}_{\rm L}(k)$,
and by both loops and long strings ${\cal P}_{\rm tot}(k)$.
The dynamic range here extends from  $0.6$ to $7.5\eta_{\rm eq}$.
As expected
${\cal P}_{\rm L}(k)$
scales more moderately than $f^2$
but more strongly than $f$ (see (\ref{rhol})).
It is also apparent that
the perturbations induced by long strings and by loops
are positively correlated with
${\cal P}_{\rm tot}(k) > {\cal P}_{\rm L}(k)+{\cal P}_\infty (k)$
throughout the whole scale range.
This positive correlation between loops and long strings
boosts the large-scale ${\cal P}_\infty (k)$ by a factor of
$1.5$, $1.8$ and $2.2$ to reach ${\cal P}_{\rm tot}(k)$
for $f=0.5, 1$ and $2$ respectively,
even if ${\cal P}_{\rm L}(k)$ is a relatively small
fraction of ${\cal P}_\infty (k)$ on these scales.

Figure \ref{figure5} shows the correlation coefficient
between the long-string and loop induced perturbations.
We see that
long strings and loops are strongly positively correlated on large scales,
but weakly correlated on small scales
where the loops dominate the perturbations (also see figure \ref{figure3}).
The threshold $k_{\rm t}$ between these two regimes
must be significantly larger than $k_{\rm L}$
because, for $k < k_{\rm L}$, ${\cal P}_{\rm L}(k)$ is well below
and roughly parallel to ${\cal P}_\infty (k)$
(see figures \ref{figure2} and \ref{figure3}).
We also verify that ${\cal P}_{\rm tot}(k)/{\cal P}_\infty (k)$ is approximately
a constant for $k < k_{\rm L}\lsim k_{\rm t}$, which
 again provides strong evidence for the fact that loops behave 
as part of the long-string
network on large scales.

\begin{figure}
  \centering 
  \leavevmode\epsfxsize=8cm \epsfbox{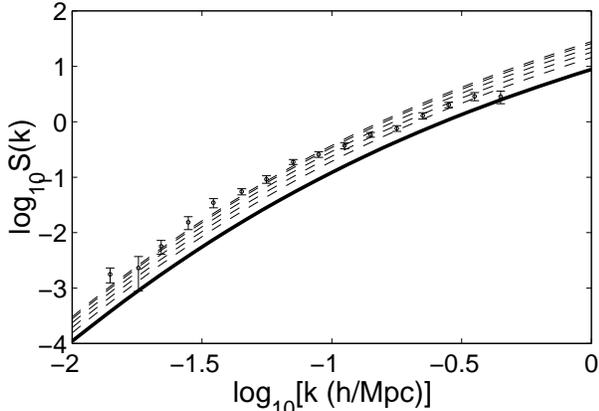}\\ 
  \caption[]
  {Comparison of the observational power spectrum \cite{PD} with
    ${\cal P}_\infty(k)$ (solid), and
    ${\cal P}_{\rm tot}(k)$ for $f=0.5,1,2,4,6$ (dashed, upwards),
    with a full dynamic range.
    }
  \label{figure4}
\end{figure}
Given these properties of the string power spectra, one can easily 
construct a semi-analytic model for
${\cal P}_{\rm tot}(k)$ as for ${\cal P}_\infty (k)$ \cite{AveShe2,AveShe3}.
We first multiply the structure function
${\cal F}(k,\eta)$ of ${\cal P}_\infty (k)$ by ${\cal J}(\eta, f)$
to account for the boost ${\cal P}_{\rm tot}(k)/{\cal P}_\infty (k)$
on large scales ($k < k_{\rm t}$),
and then by a numerically verified form
${\cal H}(k,\eta,f)=[1+(k/k_{\rm t})^4]^{1/4}$
to account for the turnover for $k>k_{\rm t}(\eta,f)$.
${\cal J}(\eta, f)$ is calibrated phenomenologically from simulations
deep in the
radiation era through to those deep in the matter era.
In the pseudo-scaling regime for the loop size,
$k_{\rm t}$ is revealed to be at least $10k_\xi\approx 200/\eta$
depending on $f$.
Thus we can carry out a full-dynamic-range integration to obtain
${\cal P}_{\rm tot}(k)$.
In figure \ref{figure4}
we compare this ${\cal P}_{\rm tot}(k)$ and ${\cal P}_\infty(k)$
\cite{AveShe2,AveShe3} with observations \cite{PD}.
The background cosmology is $\Omega_{\rm c}=0.15$, $\Omega_\Lambda=0.85$ and
$h=0.7$, and
we have used the COBE normalization
$G \mu =1.7 \times 10^{-6}$ \cite{ACDKSS} throughout.
Since loops are point-like and they have little impact through the
Kaiser-Stebbins effect on COBE-scale CMB anisotropies,
we expect this normalization to be very weakly dependent on the value of $f$;
indeed, loops were found to be negligible in ref.~\cite{ACSSV}.
Thus we see from figure \ref{figure4} that
for $f \gsim 0.5$, loops can contribute significantly to the total power 
spectrum and ease the large-scale bias problem seen previously
\cite{AveShe2,AveShe3,against}.
Definite conclusions, therefore, 
about biasing in cosmic string models will need further advances in 
determining the magnitude of the parameter $f$, while all future large-scale
structure simulations will now require the 
the inclusion of loops.

These additional complications in modelling cosmic string structure 
formation are most obvious on small scales, where even higher resolution 
and large dynamic range simulations will be required.  However, 
we expect the power spectrum on large scales to be only weakly dependent on
the details of loop formation.
Within the present pseudo-scaling regime for loop size,
we know that $k_{\rm t}\gsim k_{\rm L} \gsim 10k_\xi$
as shown in figure~\ref{figure2} and discussed previously.
Taking this extreme minimum $k_{\rm t}=10k_\xi$, then, we find that
the semi-analytic model over the full dynamic range gives
at most a $2\%$ difference in ${\cal P}_{\rm tot}(k)$ 
for $k < 1h{\rm Mpc}^{-1}$ when the filament term ${\cal H}(k,\eta,f)$ is 
excluded from 
${\cal F}(k,\eta)$
(for $\Omega_{\rm c}=0.15$, $\Omega_\Lambda=0.85$ and $h=0.7$).
This means that 
although the simulations described in this letter are already on 
the verge of present computer capabilities,
a further detailed study on small scales will improve only the overall
normalization of ${\cal P}_{\rm tot}(k)$
but not the shape revealed here, which should be a robust feature.
We note that advances in understanding loop formation mechanisms 
will also be crucial in quantifying the importance of 
the gravitational radiation background 
emitted by cosmic string network and its effect on large-scale 
structure and CMB anisotropies\cite{AveCal2}.


\subsection{\bf Conclusion}
\label{conc}

In this Letter we have described the results of high-resolution
numerical simulations of structure formation
seeded by a cosmic string network with a large dynamic range,
taking into account for the first time the loops produced by
the network.
We show that on large scales the loops behave 
like part of the long-string network and
can therefore contribute significantly to the 
total power spectrum of density perturbations,
provided their lifetime is not much smaller than one Hubble time.
At present,
the typical size and lifetime of loops formed by a string network 
remains to be studied in more detail; the problem is both computationally 
and analytically challenging.
However, within the scale range of interest further developments in this area
have the potential to affect the overall amplitude of the spectrum,
while leaving the shape largely unchanged.
The results presented here provide further encouragement for more detailed
work on both the nature of cosmic string evolution and the large-scale
structures they induce in cosmologies with 
$\Gamma\,=\,\Omega h\,=\,0.1$--$0.2$.


\acknowledgements

We would like to thank Carlos Martins for useful conversations.
P.\ P.\ A.\ is funded by JNICT (Portugal) under the `Program PRAXIS XXI'
(PRAXIS XXI/BPD/9901/96).
J.\ H.\ P.\ W.\ is funded by CVCP (UK) under the `ORS scheme'
(ORS/96009158) and by Cambridge Overseas Trust (UK).
B.\ A.\ acknowledges support from NSF grant PHY95-07740.
This work was performed on COSMOS, the Origin2000 owned by the UK
Computational Cosmology Consortium, supported by Silicon Graphics/Cray
Research, HEFCE and PPARC.




\end{document}